\documentclass[aps,twocolumn,groupedaddress,showkeys,showpacs]{revtex4}

\begin{document}
\bibliographystyle{apsrev}


\title{Origin of Light}


\author{Xiao-Gang Wen}
\homepage{http://dao.mit.edu/~wen}
\affiliation{Department of Physics, Massachusetts Institute of Technology,
Cambridge, Massachusetts 02139
}


\date{July, 2001}

\begin{abstract}
The existence of light (a \emph{massless} $U(1)$ gauge boson) is one of
unresolved mysteries in nature.
In this paper, we would like to propose that light is
originated from certain quantum orders in our vacuum.  We will
construct quantum models on lattice to demonstrate that some quantum orders
can give rise to light without breaking any symmetries and without any fine
tuning.  Through our models, we show that the existence of light can simply be
a phenomenon of quantum coherence in a system with many degrees of freedom.
Massless gauge fluctuations appears commonly and naturally in strongly
correlated systems.
\end{abstract}
\pacs{11.15.-q}
\keywords{Quantum orders, Gauge theory}

\maketitle


In an attempt to explain the meaning of ``empty space'' to a young child, I
said ``space is something not made of atoms.'' He replied ``Then you were
wrong to tell me last time that only light is not made of atoms.'' Indeed,
light and gravity are two singular forms of ``matter'' which are very
different from other  forms of matter such as atoms, electrons, \etc.
(Here I assume space = gravity.) The existences of light and gravity -- two
massless gauge bosons -- are two big mysteries in nature.

Massless particles are very rare in nature. In fact photon and graviton are
the only two massless particles known to exist. In condensed matter
systems, one encounters more kinds of gapless excitations. However, with a
few exceptions, all the gapless excitations exist 
because the ground state of the system has a special property called
spontaneous breaking of a continuous symmetry.\cite{N6080,G6154}
For example, gapless phonons exist in a
solid because a solid break the continuous translation symmetries. There
are precisely three kinds of gapless phonons since the solid breaks three
translation symmetries in $x$, $y$ and $z$ directions. Thus we
can say that the origin of gapless phonons is the translation symmetry
breaking in solids.

With the above understanding of the origin of gapless phonon in solids, we
would like to ask what is the origin of light?  Here we will adopt a point
of view that all particles, such as photons, electrons, \etc, are
excitations above a ground state -- the vacuum.  The properties of those
particles reflect the properties of the vacuum.  With this point of view,
the question on the origin of light become a question on the properties of
vacuum that allow and protect the existence of light.

If light behaved like phonons in solids, then we could conclude that our
vacuum break a continuous symmetry and light would be originated from
symmetry breaking. However, in reality, light does not behave like the
phonons. In fact there are no phonon-like particles (or more precisely,
massless Nambu-Goldstone bosons) in nature. From the lack of massless
Nambu-Goldstone
bosons, we can conclude that there is no continuous symmetry breaking in
our vacuum. If the vacuum does not break any continuous symmetry, then what
makes light to exist?

In a recent work,\cite{Wqos,Wqo} a concept -- quantum order -- was introduced to
describe a new kind of orders that generally appear in quantum states at
zero temperature. Quantum orders that characterize universality classes of
quantum states (described by \emph{complex} ground state wave functions) is
much richer then classical orders that characterize universality classes of
finite temperature classical states (described by \emph{ positive}
probability distribution functions).  In contrast to classical orders,
quantum orders cannot be described by broken symmetries and the associated
order parameters. A new mathematical object -- projective symmetry group
(PSG) -- was introduced to characterize quantum orders.  In a sense,
we can view a quantum order as a dancing pattern in which particles
waltz around each other in a ground state. The PSG is a mathematical
description of the dancing pattern. In contract, the classical order in a
crystal just describes a static positional pattern, which can be characterized
by symmetries.

In \Ref{Wqo}, various quantum orders are studied. It was found that different
quantum orders (characterized by different PSG's) can have distinct low energy
properties. In particular, certain quantum orders allow and protect gapless
excitations even without breaking any continuous symmetry. This leads us to
propose that it is the quantum order in our vacuum that allow and protect the
existence of light.  In another word, light is originated from quantum order.

To support our idea, in the following, we are going to study a concrete
$SU(N_f)$ spin model\cite{AM8874,RS9068} in 3D, and show that its ground state
contains a gapless $U(1)$ gauge fluctuation. More importantly, we will
identify the quantum order (or the PSG) in the ground state and argue that the
gapless property of the $U(1)$ gauge fluctuations is a robust property
protected by the quantum orders. A small change of the Hamiltonian cannot
destroy the gapless $U(1)$ gauge fluctuations.  We would like to mention that
a connection between QCD and a lattice spin model was pointed out in
\Ref{L98}, using concept of quantum critical point.  In our example, we will
see that the massless property of light is not due to criticality. It is a
generic property of a quantum phase.



We start with a $SU(N_f)$-spin model\cite{AM8874,RS9068} on a 3D cubic lattice.
The states on each site form a representation of rank $N_f/2$ antisymmetric
tensor of $SU(N_f)$. 
We note that those states can be viewed as states of $N_f/2$ fermions
with fermions $\psi_{a\v i}$, $a=1,...,N_f$ 
in the fundamental representation of $SU(N_f)$.  
Thus we can write down the Hamiltonian of our model in terms of 
the fermion operators:
\begin{align}
\label{SUNfHam}
 H = J_P \sum_{\vev{\v i_1\v i_2\v i_3\v i_4}} \left(
S^{ab}_{\v i_1} S^{bc}_{\v i_2} S^{cd}_{\v i_3} S^{da}_{\v i_4}
+ h.c. \right)
\end{align}
where the sum is over all plaquettes $\vev{\v i_1\v i_2\v i_3\v i_4}$,
\begin{align}
\label{parton}
 S^{ab}_{\v i} = \psi^\dag_{a \v i} \psi_{b \v i} - N_f^{-1}\del_{ab}
 \psi^\dag_{c \v i} \psi_{c \v i}  .
\end{align}
The Hamiltonian
has three translation symmetries and six parity symmetries $P_x: x\to -x$,
$P_y: y\to -y$, $P_z: z\to -z$, $P_{xy}: x\leftrightarrow y$, $P_{yz}:
y\leftrightarrow z$, $P_{zx}: z\leftrightarrow x$.  The Hamiltonian also
has a charge conjugation symmetry $C: \psi_{a\v i}\to \psi^\dag_{a\v i}$. 

To find the ground state of the above systems,
we will use the projective construction
(which is a generalization of slave-boson 
approach\cite{BZA8773,BA8880,AM8874}) 
to construct the
ground state. We start with a mean-field parton Hamiltonian
\begin{align}
 H_{mean}=& -\sum_{\<\v i\v j\>} \left( 
\psi^\dag_{a,\v i} \chi_{\v i\v j} \psi_{a,\v j} + h.c. 
\right)
\end{align}
where
$ \chi_{\v i\v j}^\dag = \chi_{\v j\v i} $.
The mean-field Hamiltonian allows us to construct a trial wave function
for the ground state of the $SU(N_f)$-spin system \Eq{SUNfHam}:
\begin{align}
 |\Psi_{trial}^{(\chi_{\v i\v j})}\> = \cP |\Phi_{mean}^{(\chi_{\v i\v j})}\>
\end{align}
where $|\Phi_{mean}^{(\chi_{\v i\v j})}\>$ is the ground state of the
mean-field Hamiltonian $H_{mean}$ and $\cP$ is the projection to states
with $N_f/2$ fermion per site. Clearly the mean-field ground state is a
functional of $\chi_{\v i\v j}$. The proper values of $\chi_{\v i\v j}$
are obtained by minimizing the trial energy
$E= \< \Psi_{trial}^{(\chi_{\v i\v j})}| H |\Psi_{trial}^{(\chi_{\v i\v j})}\>$.

The relation between the physical operator $S^{ab}$ and the parton operator
$\psi_a$ essentially defines the projective construction.\cite{Wpc} For
example the fact that the operator $S^{ab}_{\v i}$ is invariant under local
$U(1)$ transformations
\begin{align}
 \psi_{a \v i} \to & e^{i\th_{\v i}} \psi_{a \v i}, &
 S^{ab}_{\v i} \to & S^{ab}_{\v i} 
\end{align}
determines the high energy $U(1)$ gauge structure in the parton mean-field
theory:
\begin{align}
\label{U1trn}
 \psi_{a \v i} \to & e^{i\th_{\v i}} \psi_{a \v i}, &
\chi_{\v i\v j} \to & e^{i\th_{\v i}} \chi_{\v i\v j}e^{-i\th_{\v i}} 
\end{align}
The $U(1)$ gauge structure has a very real meaning:
two gauge equivalent ansatz give rise to the same physical state
after projection
\begin{align}
\label{gauge}
 |\Psi_{trial}^{(\chi_{\v i\v j})}\>
 = |\Psi_{trial}^{(e^{i\th_{\v i}}\chi_{\v i\v j}e^{-i\th_{\v j}})}\>
\end{align}

Usually it is hard to calculate the trial energy
$E= \< \Psi_{trial}| H |\Psi_{trial}\>$. In the following, we will calculate
$\chi_{\v i\v j}$ by minimizing the mean-field energy
$E_{mean}=\<\Phi_{mean}^{(\chi_{\v i\v j})}|H
          |\Phi_{mean}^{(\chi_{\v i\v j})}\>$
which approaches to the exact ground state energy in the large $N_f$ 
limit.\cite{AM8874,RS9068} We assume
$|\Phi_{mean}^{(\chi_{\v i\v j})}\>$ 
to respect the $SU(N_f)$ symmetry, which leads to
$ \< \Phi_{mean}^{(\chi_{\v i\v j})}|\psi_{a\v i}\psi^\dag_{b\v j}
 |\Phi_{mean}^{(\chi_{\v i\v j})}\>
 =\del_{ab} \t\chi_{\v i\v j} $.
We find
\begin{align}
\frac{E_{mean}}{  J_P N_f^4}  =
 \sum_{\vev{\v i_1\v i_2\v i_3\v i_4}} \left(
\t\chi_{\v i_1\v i_2}\t\chi_{\v i_2\v i_3}
\t\chi_{\v i_3\v i_4}\t\chi_{\v i_4\v i_1}
+h.c. \right) 
+ O(N_f^3)  \nonumber 
\end{align}
Since a $\pi$-flux in a plaquette make $\t\chi_{\v i_1\v i_2}\t\chi_{\v
i_2\v i_3} \t\chi_{\v i_3\v i_4}\t\chi_{\v i_4\v i_1}$ to be a negative
number, we expect the ansatz that minimize $E_{mean}$ to have $\pi$
flux on every plaquette.
Such an ansatz can be constructed and takes a form
\begin{align}
\label{U1a}
\bar \chi_{\v i,\v i+\hat{\v x}} =& -i\chi , &
\bar \chi_{\v i,\v i+\hat{\v y}} =& -i(-)^{i_x}\chi ,  \nonumber\\
\bar \chi_{\v i,\v i+\hat{\v z}} =& -i(-)^{i_x+i_y}\chi . 
\end{align}
Such an ansatz, after projection, gives rise to a correlated ground state
for our $SU(N_f)$-spin system.

In the momentum space, the mean-field Hamiltonian has a form
\begin{align}
 H_{\text{mean}}=& -{\sum_{\v k}}^\prime 
\Psi^\dag_{a,\v k} \Ga(\v k) \Psi_{a,\v k}
\end{align}
where
\begin{align}
\Psi_{a,\v k}^T =& (
\psi_{a,\v k},
\psi_{a,\v k+\v Q_x},
\psi_{a,\v k+\v Q_y},
\psi_{a,\v k+\v Q_x+\v Q_y} 
),  \nonumber\\
\v Q_x = & (\pi,0,0),\ \ \ \v Q_y =  (0,\pi,0), \nonumber\\
\Ga(\v k) =& 
2\chi( \sin(k_x) \Ga_1 + \sin(k_y) \Ga_2 + \sin(k_z) \Ga_3 ) \nonumber 
\end{align}
and
$\Ga_1 = \tau^3\otimes \tau^0$,
$\Ga_2 = \tau^1\otimes \tau^3$, and
$\Ga_3 = \tau^1\otimes \tau^1$.
The momentum summation is over a range $k_x \in (-\pi/2,\pi/2)$, $k_y \in
(-\pi/2,\pi/2)$, and $k_z \in (-\pi,\pi)$.  Since $\{ \Ga_i, \Ga_j \}
=2\del_{ij}$, $i,j=1,2,3$, we find partons have a dispersion
\begin{align} 
E(\v k) = \pm 2\chi \sqrt{ \sin^2(k_x) + \sin^2(k_y) + \sin^2(k_z)} 
\end{align} 
The mean-field ground state $|\Phi_{mean}\>$ is obtained by filling the
negative energy branch.  We see that the dispersion has two nodes at $\v k =
0$ and $\v k = (0,0,\pi)$. Thus there are $2N_f$ massless four-component
Dirac fermions in the continuum limit. The low energy theory has Lorentz
symmetry.  Including the collective phase fluctuations of the ansatz, 
the low energy effective theory has a form
\begin{align}
 L = \sum_{\v i} 
\psi^\dag_{a,\v i} i (\prt_t + i a_0) \psi_{a,\v j} 
  +\sum_{\v i\v j}  
\psi^\dag_{a,\v i} \bar \chi_{\v i\v j}e^{ia_{\v i\v j}} \psi_{a,\v j} 
\nonumber 
\end{align}
In the continuum limit, it becomes
$ \cL = \bar \psi_{\al a} D_\mu \ga^\mu \psi_{\al a}$
with $D_\mu =\prt_\mu + i a_\mu$, $\al =1,2$, and $\ga_\mu$ are $4\times 4$
Dirac matrices. Integrating out the high energy fermions will generate
dynamics for the $a_\mu$ field (see \Eq{effL}).
We see that our correlated ground state,
$\cP |\Phi_{mean}^{(\bar \chi_{\v i\v j})}\>$, support a massless $U(1)$ gauge
fluctuations and $2N_f$ massless Dirac fermions.

In the following we would like to argue that the appearance of the massless
$U(1)$ gauge fluctuations and the massless Dirac fermions is not a special
property of the particular state constructed above. It is a universal property
of a quantum phase (characterized by a particular quantum order).  We will
first find the PSG of the constructed state. Then we will argue that the PSG
is a universal property of a quantum phase by showing that radiative
corrections cannot change the PSG. Last we will show that any state described
by the same PSG (\ie any state in the same quantum phase) will have the same
massless $U(1)$ gauge fluctuations and the same massless Dirac fermions.  We
would like to remark that the stability of the massless $U(1)$ gauge
fluctuations in 3+1D is not new. But the stability of massless Dirac
fermions may be new and the PSG approach put the stability of the
massless $U(1)$ gauge fluctuations and the stability of massless Dirac
fermions on the same footing.

The PSG\cite{Wqo} that characterizes the quantum order in the above correlated
state is given by
\begin{align}
\label{U1PSG}
 G_x(\v i) =& (-)^{i_y+i_z} e^{i \th_x} &
 G_y(\v i) =& (-)^{i_z} e^{i \th_y} \nonumber\\
 G_z(\v i) =& e^{i \th_z} &
 G_{px}(\v i) =& (-)^{i_x} e^{i \th_{px}} \nonumber\\
 G_{py}(\v i) =& (-)^{i_y} e^{i \th_{py}} &
 G_{pz}(\v i) =& (-)^{i_z} e^{i \th_{pz}} \\
 G_{pxy}(\v i) =& (-)^{i_xi_y} e^{i \th_{pxy}} &
 G_{pyz}(\v i) =& (-)^{i_yi_z} e^{i \th_{pyz}} \nonumber\\
 G_C(\v i) =& (-)^{\v i} e^{i \th_t} &
 G_{pzx}(\v i) =& (-)^{(i_x+i_y)(i_y+i_z)} e^{i \th_{pzx}} \nonumber 
\end{align}
The invariant gauge group (IGG) of the ansatz is $\cG = \{ e^{i\th}
\} = U(1)$, which is a (normal) subgroup of the PSG. 
Here $G_{x,y,z}$ are the gauge transformations associated with the three
translations, $G_{px,py,pz}$ are associated with the three parities $P_x$,
$P_y$, $P_z$, and $G_{pxy,pyz,pzx}$ are associated with the other three
parities $P_{xy}$, $P_{yz}$, $P_{zx}$, and $G_C$ is associated with charge
conjugation transformation $C: \chi_{\v i\v j}\to -\chi_{\v i\v j}$. 
The ansatz is invariant, say, under the party transformation $P_x$ followed by
the gauge transformation $G_{px}$.

To show that the PSG is a universal property of a quantum phase,\cite{Wqo} 
we start with the mean-field state
characterized by $\chi_{\v i\v j} = N_f^{-1}\<\psi_{a \v i}\psi^\dag_{a \v
j}\>$.  If we include perturbative fluctuations around the mean-field state,
we expect $\chi_{\v i\v j}$ to receive radiative corrections $\del \chi_{\v
i\v j}$.  However, the perturbative fluctuations can only change $\chi_{\v i\v
j}$ in such a way that $\chi_{\v i\v j}$ and $\chi_{\v i\v j}+\del \chi_{\v
i\v j}$ have the same projective symmetry group.  This is because if $\chi_{\v
i\v j}$ and the Hamiltonian have a symmetry, then $\del \chi_{\v i\v j}$
generated by perturbative fluctuations will have the same symmetry.  The
transformation generated by an element in PSG just behave like a symmetry
transformation in the perturbative calculation. The mean-field ground state
and the mean-field Hamiltonian are invariant under the transformations in the
PSG. Therefore, $\del \chi_{\v i\v j}$ generated by perturbative fluctuations
will also be invariant under the transformations in the PSG.  Thus the
perturbative fluctuations cannot change the PSG of an ansatz.  Also if we
perturb the $SU(N_f)$-spin Hamiltonian \Eq{SUNfHam} without breaking any
symmetries, the induced $\del \chi_{\v i\v j}$ is still invariant under the
transformations in the PSG.  Thus the PSG is robust
against small perturbations of the Hamiltonian and it is a universal property
of a quantum phase. The PSG can change only when the fluctuations have an
infrared divergence which will drive a phase transition.

To understand 
how quantum orders and PSG's protect the gapless excitations
without breaking any symmetries,
we would like to first find out the possible fluctuations at low energies.
The first kind of low energy excitations are described by the
particle-hole excitations of the fermions across the Fermi points.
The $SU(N_f)$-spin wave functions for such kind of excitations are given by
$ |\Psi_{exc}^{(\bar\chi_{\v i\v j})}\> 
= \cP \psi^\dag_{\v k_1} \psi_{\v k_2}|\Phi_{mean}^{(\bar\chi_{\v i\v j})}\>$.
The second kind of low energy excitations are the collective excitations
described by the phase fluctuations of the ansatz:
$\chi_{\v i\v j}=\bar\chi_{\v i\v j}e^{ia_{\v i\v j}}$.
The $SU(N_f)$-spin wave-functions for such collective excitations are given by
$ |\Psi_{exc}^{(\bar\chi_{\v i\v j}e^{ia_{\v i\v j}})}\> $.

To see that the massless fermion excitations are
protected by the quantum order, we need
to consider the most generic ansatz $\chi_{\v i\v j}$ that have the same
PSG \Eq{U1PSG} and check if the fermions are still massless for those
generic ansatz. The most general translation symmetric ansatz has a form
\begin{align}
 \chi_{\v i,\v i+\v m} =& \chi_{\v m} (-)^{i_ym_z} (-)^{i_x(m_y+m_z)}
\end{align}
To have the parity symmetry $\v i \to -\v i$, the ansatz should be
invariant under transformation $\v i\to -\v i$ followed by a gauge
transformation $(-)^{\v i}$. This requires that
$ \chi_{\v m} = (-)^{\v m} \chi_{-\v m}  = (-)^{\v m} \chi_{\v m}^\dag $.
To have charge conjugation symmetry $\chi_{\v i\v j}$ must change sign
under gauge transformation $W_{\v i}=(-)^{\v i}$. This requires that
$ \chi_{\v m} = 0$, if $\v m = even$.
Thus the most general ansatz has a form
\begin{align}
 \chi_{\v i,\v i+\v m} =& \chi_{\v m} (-)^{i_ym_x} (-)^{i_x(m_x+m_y)}
 \nonumber\\
 \chi_{\v m} = & 0,\ \ \ \ \hbox{if $\v m = even$}  \nonumber\\
 \chi_{\v m} = & - \chi_{\v m}^\dag = -\chi_{-\v m}
\end{align}
In the momentum space, $\chi$ vanishes at $\v k = 0$ and $(0,0,\pi)$.  Thus
the PSG protect the massless Dirac fermions.

To see that the massless collective fluctuations described by $a_{\v i\v j}$
are protected by the quantum order, we need to show the collective
fluctuations are massless for the most general ansatz that have the same PSG
\Eq{U1PSG}.  For any ansatz that is invariant on the PSG, it is also invariant
under the IGG $\cG = \{ e^{i\th} \} = U(1)$ which is subgroup of the PSG.  In
this case $a_{\v i\v j}$ and $\t a_{\v i\v j}=a_{\v i\v j}+\th_{\v i} -\th_{\v
j}$ label the same quantum state (and are said to be gauge equivalent). (See
\Eq{gauge}.) We see that $a_{\v i\v j}$ describes a $U(1)$ gauge fluctuation.
Since the energy of the fluctuation $E(a_{\v i\v j})$ satisfies $E(a_{\v i\v
j}) = E(\t a_{\v i\v j})$, the mass term $(a_{\v i\v j})^2$ is not allowed and
there is no Anderson-Higgs mechanism to give $U(1)$ gauge field a mass. Thus
the $U(1)$ gauge fluctuations are gapless for any ansatz that has the PSG
\Eq{U1PSG}.

In the standard analysis of the stability of the massless excitations, one
needs to include all the counter terms that have the right symmetries into the
Lagrangian, since those terms can be generated by perturbative fluctuations.
Then we examine if those allowed counter terms can destroy the massless
excitations or not.  In our problem, we need to consider all the possible
corrections to the mean-field ansatz.  However, the new feature here is that
it is incorrect to use the symmetry group to determine the allowed
corrections.  We should use PSG to determine the allowed corrections in our
analysis of the stability of the massless excitations.

Next we consider a lattice model that contains both massless and massive
fermions.  The mean-field Hamiltonian is given by
\begin{align}
 H_{mean}=& 
 -\sum_{\<\v i\v j\>} 
\left( 
\psi^\dag_{a,\v i} \chi_{\v i\v j} \psi_{a,\v j} + h.c. 
\right) \\
  & -\sum_{\<\v i\v j\>} 
\left( 
\la^\dag_{\al,\v i} \chi_{\v i\v j} \tau^3 \la_{\al,\v j} + h.c. 
\right)
 -\sum_{\v i} 
\la^\dag_{\al,\v i} m \tau^1 \la_{\al,\v i}  \nonumber 
\end{align}
where $a=1,...,N_f$, $\al=1,...,N_f^\prime$, $\la_{\al}$ is a doublet:
$\la_\al^T =( \la_\al^{(1)}, \la_\al^{(2)})$, and $\chi_{\v i\v j}$ is
given in \Eq{U1a}.  The model has a $U(1)$ gauge structure defined by the
gauge transformation
$ \psi_{a,\v i}\to  e^{i\th_{\v i}} \psi_{a,\v i}$,
$\la_{a,\v i}\to e^{i\th_{\v i}} \la_{a,\v i}$,  and
$\chi_{\v i\v j} \to  e^{i(\th_{\v i}-\th_{\v j})} \chi_{\v i\v j}$.
Clearly, the model has a $SU(N_f)\times SU(N_f^\prime)$ global symmetry.
The gauge invariant physical operators are given by
$\psi^\dag_{a,\v i}\psi_{b,\v i}$,
$\la^\dag_{a',\v i}\la_{b',\v i}$,
and $\psi^\dag_{a,\v i}\la_{a',\v i}$
In the momentum space, the above mean-field Hamiltonian becomes
\begin{align}
 H_{\text{mean}}=& -{\sum_{\v k}}^\prime 
\Psi^\dag_{a,\v k} \Ga(\v k) \Psi_{a,\v k}
+\La^\dag_{a,\v k} \t\Ga(\v k) \La_{a,\v k}
\end{align}
where
$\La_{a,\v k}^T = (
\la_{a,\v k},
\la_{a,\v k+\v Q_x},
\la_{a,\v k+\v Q_y},
\la_{a,\v k+\v Q_x+\v Q_y} 
)$,
$\v Q_x =  (\pi,0,0),\ \ \ \v Q_y =  (0,\pi,0)$,
$\t\Ga(\v k) = 
2\chi( \sin(k_x) \t\Ga_1 + \sin(k_y) \t\Ga_2 + \sin(k_z) \t\Ga_3 ) 
+m \t\Ga_m$,
and
$\t\Ga_1 = \tau^3\otimes \tau^0\otimes \tau^3$,  
$\t\Ga_2 = \tau^1\otimes \tau^3\otimes \tau^3$,
$\t\Ga_3 = \tau^1\otimes \tau^1\otimes \tau^3$, and
$\t\Ga_m = \tau^0\otimes \tau^0\otimes \tau^1$.
We see that there are $2N_f$ massless Dirac fermions and $4N_f^\prime$
massive Dirac fermions in the continuum limit.  Those fermions carry
crystal momenta near $\v k=0$ and $\v k = (0,0,\pi)$.  The PSG that
characterizes the above mean-field state is still given by \Eq{U1PSG}, which
acts on both $\psi$ and $\la$. Since IGG = $U(1)$, the fluctuations around
the mean-field state contain a $U(1)$ gauge field at low energies. After
including the $U(1)$ gauge field and in the continuum limit, the low energy
effective theory takes a form
\begin{align}
\cL = \sum_{I=1}^{2N_f}\bar \psi_I D_\mu \ga^\mu \psi_I + 
\sum_{J=1}^{4N_f^\prime}\bar \la_J D_\mu \ga^\mu \la_J + m \bar \la_J \la_J
\end{align}
where $D_\mu = \prt_\mu +i a_\mu$ and $\ga^\mu$ are the $\ga$-matrices.
After integrating out high energy fermions, we get
\begin{align}
\label{effL}
\cL = \sum_{I=1}^{2N_f}\bar \psi_I D_\mu \ga^\mu \psi_I + 
\frac{\al^{-1}}{8\pi}(\v E^2-\v B^2)
\end{align}
where the fine structure constant at energy scale $E$ is
\begin{equation}
\label{alpha}
 \al^{-1}(E)  = \frac{2}{3\pi}[2N_f\ln(E_0/E) +4 N_f'\ln(E_0/m)]
\end{equation}
where $E_0$ is the lattice energy scale. We have assumed $m\ll E_0$.

In this paper we propose that light is originated from the quantum order in
our vacuum. To demonstrate this idea, we construct a lattice model with
$SU(N_f)\times SU(N_f')$ spins. We show that in the large $N_f$ and $N_f'$
limit, our lattice model has a ground state characterized by
the quantum order \Eq{U1PSG}. We find that the PSG (or the quantum
order) protects the gapless $U(1)$ gauge fluctuations and the
massless Dirac fermions (when $N_f>0$). 
We note that the low energy fermion excitations in our model have the Lorentz
invariance, which is also protected by the quantum order.
Certainly our lattice ground state
does not describe the quantum order of the real vacuum.
It would be interesting to find a lattice model that gives rise to
$U(1)\times SU(2)\times SU(3)$ low energy gauge structure together with
chiral leptons and quarks.


%



This research is
supported by NSF Grant No. DMR--97--14198.

\bibliography{/home/wen/bib/wencross,/home/wen/bib/wen,/home/wen/bib/htc,/home/wen/bib/part}

\end{document}